\documentclass[aps,prd,showpacs,floatfix,nofootinbib,12pt]{revtex4}

\usepackage{color,amsmath,amssymb,graphicx,latexsym,subfigure}
\usepackage{threeparttable,txfonts}
\usepackage{hyperref}
\usepackage{ulem}

\def\ga   {\gamma}

\def\sig   {\sigma}

\def\nn{\nonumber}

\def\mof {\mathcal{O}^{(5)}}
\def\mos {\mathcal{O}^{(6)}}

\newcommand{\gev}{\ensuremath{\,\mathrm{GeV}}}

\newcommand{\mev}{\ensuremath{\,\mathrm{MeV}}}
\newcommand{\mchi}{\ensuremath{m_\chi}}
\newcommand{\sv}{\ensuremath{\langle\sigma v\rangle}}

\begin{document}

{\small
\begin{flushright}
CP3-Origins-2018-020 DNRF90
\end{flushright} }

\voffset 1.25cm

\title{Consistency test of the AMS-02 antiproton excess with direct 
detection data based on the effective field theory approach}

\author{Ming-Yang Cui$^{a,b}$\footnote{mycui@pmo.ac.cn}}
\author{Wei-Chih Huang$^c$\footnote{huang@cp3.sdu.dk}}
\author{Yue-Lin Sming Tsai$^{d}$\footnote{smingtsai@gate.sinica.edu.tw}}
\author{Qiang Yuan$^{a,e,f}$\footnote{yuanq@pmo.ac.cn}}

\affiliation{
$^a$Key Laboratory of Dark Matter and Space Astronomy, Purple
Mountain Observatory, Chinese Academy of Sciences, Nanjing
210008, China \\
$^b$School of Physics, Nanjing University, Nanjing 210093, China \\
$^c$CP$^3$ Origins, University of Southern Denmark, Campusvej 55, DK-5230 Odense M, Denmark\\
$^d$Institute of Physics, Academia Sinica, Nangang, Taipei 11529, Taiwan \\
$^e$School of Astronomy and Space Science, University of Science and 
Technology of China, Hefei, Anhui 230026, China \\
$^f$Center for High Energy Physics, Peking University, Beijing 100871, China
}

\begin{abstract}
The potential antiproton excess in the AMS-02 data is of much interest
and can probably come from dark matter annihilations. Based on the 
effective field theory approach, in this work we investigate the compatibility of the 
DM interpretation of the AMS-02 antiproton excess and the null results from 
direct detection experiments, LUX, PandaX-II, and XENON1T. 
We focus on dimension-five and -six operators with fermion DM.
Only one of dimension-five and one of dimension-six operators can 
successfully account for the antiproton excess, while the rest either
are excluded by direct detection or require very small cut-off scales
which invalidate the effective field theory approach.

\end{abstract}

\date{\today}

\pacs{95.35.+d,96.50.S-}

\maketitle

\section{Introduction}

In the past decade, there has been significant progress in dark matter
(DM) searches, either from underground nuclear recoil experiments
(direct detection; DD) or from space-borne $\gamma$-ray and cosmic ray 
observations known as indirect detection~(ID). The sensitivity of DM-nucleon 
interaction measurements has been improved rapidly in the recent years. 
The liquid xenon detectors such as LUX~\cite{Akerib:2016vxi}, 
PandaX-II~\cite{Tan:2016zwf,Cui:2017nnn}, and the first ton-scale detector 
XENON1T~\cite{Aprile:2017iyp} gradually push the bounds on the DM-nucleon 
interaction strength closer to the neutrino floor, which refers to the
background from solar and atmospheric neutrinos scattering off nuclei.
The latest XENON1T result sets an upper limit on the spin-independent~(SI) 
DM-nucleon cross-section to be $\sim10^{-46}\,\rm{cm}^2$ for the DM mass 
$\mchi$ around $30\gev$. Such a stringent limit not only implies small 
DM-quark coupling strength but also constrains loop-induced 
interactions in leptophilic DM scenarios; see, for example, Refs.~\cite{Kopp:2009et,Huang:2013apa,DEramo:2017zqw,Liu:2017kmx}.
On the other hand, there exist several anomalies in the cosmic ray 
and $\gamma$-ray data, such as the Galactic center GeV excess~\cite{Hooper:2010mq,Hooper:2011ti}, the positron fraction excess
\cite{Adriani:2008zr,Aguilar:2013qda,Accardo:2014lma}, and the excess on 
the total electron and positron flux~\cite{Chang:2008aa,
Abdo:2009zk,Aguilar:2014fea,Ambrosi:2017wek}.
In addition, a potential antiproton excess around $10$ 
GeV~\cite{Cuoco:2016eej,Cui:2016ppb} has been recently suggested 
based on the very precise AMS-02 data~\cite{Aguilar:2016kjl}.
Although astrophysical sources such as pulsars can account for 
the $\gamma$-ray and positron excesses~(for a recent review, see 
Ref.~\cite{Bi:2014hpa}), it is non-trivial to simultaneously accommodate 
the antiproton excess and satisfy the constraints from the Boron-to-Carbon
(B/C) ratio~\cite{Aguilar:2016vqr} with an astrophysical model.
Alternatively, DM can also provide a possible solution to the antiproton 
excess. Besides, DM models, which can realize the AMS02 antiproton 
excess, may simultaneously explain the $\gamma$-ray excess at the Galactic 
center and the tentative $\gamma$-ray emission from a few dwarf 
galaxies or clusters~\cite{Cui:2016ppb,Cuoco:2017rxb,
Li:2015kag,Geringer-Sameth:2015lua,Liang:2016pvm,Arcadi:2017vis}.

If DM particles annihilate into standard model~(SM) final states, the 
favored DM mass $\mchi$ by the  antiproton excess is around $(60-100)~
\gev$, and the corresponding annihilation cross-section $\sv$ is, 
for example, about $(0.7-7) \times 10^{-26}\,\rm{cm}^3 s^{-1}$ for 
the $b\bar{b}$ final state~\cite{Cui:2018klo} (similar results were obtained in
Refs.~\cite{Cuoco:2017rxb,Cuoco:2017iax}). It indicates sizable couplings 
between DM and quarks, and thus one should scrutinize the DM interpretation 
with the DD data.

In this work we employ the effective field theory~(EFT) approach to 
describe the interactions between the DM and SM particles. It enables us 
to study the low-scale interactions between the DM and SM sectors in a 
generic way without considering specific underlying theories. Assuming 
DM-SM couplings of $\mathcal{O}(1)$, the cut-off scale $\Lambda$ has to 
be at least twice the mass of DM such that the underlying mediator cannot 
be on-shell, ensuring the validity of the EFT approach. On the other hand, 
the momentum exchange between DM and nuclei in DD experiments is typically 
around $100$ $\mev$, which is much lower than the DM mass and the cut-off 
scale of interest. As a result, the usage of EFT on computation of 
DM-nuclei scattering cross-sections is well-justified. 
One nonetheless has to take into account the renormalization group~(RG) 
running of the Wilson coefficients from the cut-off scale down to the 
hadronic scale~($\sim 2$ GeV), followed by the matching with the nuclear 
theory as the DM particle scatters off the whole nuclei instead of 
interacting with individual constituent quarks. To properly includes 
the effects of the RG running and non-perturbative matching, we follow
Refs.~\cite{Bishara:2016hek,Bishara:2017pfq,Bishara:2017nnn}.

This paper is organized as follows. In Sec.~\ref{sec:method} we 
describe the likelihood calculation based on the antiproton and DD data. 
In Sec.~\ref{sec:EFT}, we spell out the selective effective operators of 
dimension-five~(dim-5) and dimension-six~(dim-6), which describe how DM 
couples to the SM particles. Subsequently, we present favored regions 
on the ($\mchi$, $\Lambda$) plane by the AMS-02 antiproton data as well 
as the constraints from the combined DD data in Sec.~\ref{sec:result}. 
We summarize our study in Sec. V.

\section{Method}
\label{sec:method}

In the following, we will present our results in terms of the Bayesian posterior $95\%$ credible region (CR). 
The prior distributions employed here and how to construct the likelihood functions for the AMS02 antiproton and 
direct search data will be illustrated in this section.

\subsection{Likelihood of AMS-02 antiproton data}

The propagation of cosmic rays in the Milky Way is calculated by the
GALPROP package~\cite{Strong:1998pw,Moskalenko:1997gh}. The propagation 
and primary source parameters, denoting as background parameters 
$\boldsymbol{\theta}_{\rm bkg}$, are determined through a global 
fitting~\cite{Yuan:2018vgk} to the recent AMS-02 measurements on the 
B/C ratio~\cite{Aguilar:2016vqr} and the Carbon flux~\cite{Aguilar:2017hno}. 
The posterior distribution of $\boldsymbol{\theta}_{\rm bkg}$ from the 
fitting will be included as an updated prior. The background antiprotons 
are produced via inelastic collisions between cosmic ray nuclei and the 
interstellar medium. The recently updated parametrization of the antiproton 
production cross-section from nuclei-nuclei collisions~\cite{Winkler:2017xor} 
has been adopted to calculate the production of background antiprotons which
are then propagated with the same propagation framework. The posterior 
probability for a given antiprotons spectrum from the DM annihilation, 
$\phi_{\rm DM}(E)$, can be written as
\begin{equation}
\mathcal{P}_{\rm DM}\propto \int \mathcal{L}_{\bar{p}}
(\boldsymbol{\theta}_{\rm bkg},\kappa,\phi_{\rm DM})\,
p(\boldsymbol{\theta}_{\rm bkg})\,p(\kappa)\,
d\boldsymbol{\theta}_{\rm bkg}\,d\kappa,
\label{eq:post_dm}
\end{equation}
where $\kappa$ denotes a constant scale factor multiplied on the 
background flux which characterizes the uncertainties of the production
cross-section~\cite{Winkler:2017xor}. The symbols $p(\kappa)$ and 
$p(\boldsymbol{\theta}_{\rm bkg})$ are prior distributions of $\kappa$ 
and $\boldsymbol{\theta}_{\rm bkg}$, respectively.
The likelihood function of model parameters $(\boldsymbol{\theta}_{\rm bkg},
\kappa,\phi_{\rm DM})$ is
\begin{equation}
\mathcal{L}_{\bar{p}}(\boldsymbol{\theta}_{\rm bkg},\kappa,\phi_{\rm DM})
\propto \prod_i \exp\left[-\frac{(F_i-\kappa F_{{\rm bkg},i}-
\phi_{{\rm DM},i})^2}{2\sigma_i^2}\right] \, .
\end{equation}
In the above expression, $F_i$ and $\sigma_i$ are 
the observed flux and the corresponding error from the AMS-02 data, while $F_{{\rm bkg},i}$ and 
$\phi_{{\rm DM},i}$ correspond to the calculated background and DM-origin
antiproton flux in the $i$-th energy bin.

The propagated flux of the DM component also depends on the background 
parameters $\boldsymbol{\theta}_{\rm bkg}$, which makes the calculation 
of Eq.~(\ref{eq:post_dm}) non-trivial as shown in our previous 
work~\cite{Cui:2016ppb}. Following Ref.~\cite{Cui:2018klo} we employ an 
approximation in calculating the likelihood of the DM component as follows.
The DM-induced antiproton fluxes for different values of the background 
parameters are found to span in a not-too-wide band around the mean flux 
which is calculated with the mean parameters 
$\bar{\boldsymbol{\theta}}_{\rm bkg}$. Therefore we include an
additional constant factor $f$ to characterize the uncertainties of 
$\phi_{\rm DM}$ associated with the background parameters, i.e.,
$\phi_{\rm DM}(\boldsymbol{\theta}_{\rm bkg})=f\phi_{\rm DM}
(\bar{\boldsymbol{\theta}}_{\rm bkg})$. In this way, Eq.~(\ref{eq:post_dm}) 
can be approximated as
\begin{equation}
\mathcal{P}_{\rm DM}\propto \int \mathcal{L}_{\bar{p}}
(\boldsymbol{\theta}_{\rm bkg},\kappa,f\bar{\phi}_{\rm DM})\,
p(\boldsymbol{\theta}_{\rm bkg})\,p(\kappa)\,p(f)\,
d\boldsymbol{\theta}_{\rm bkg}\,d\kappa\,df,
\label{eq:post_dm2}
\end{equation}
in which $\bar{\phi}_{\rm DM} \equiv \phi_{\rm DM}
(\bar{\boldsymbol{\theta}}_{\rm bkg})$. We numerically confirm that the approximation
yields very similar results to those without the approximation as displayed in Appendix A.

\begin{figure}[!htb]
\includegraphics[width=0.7\textwidth]{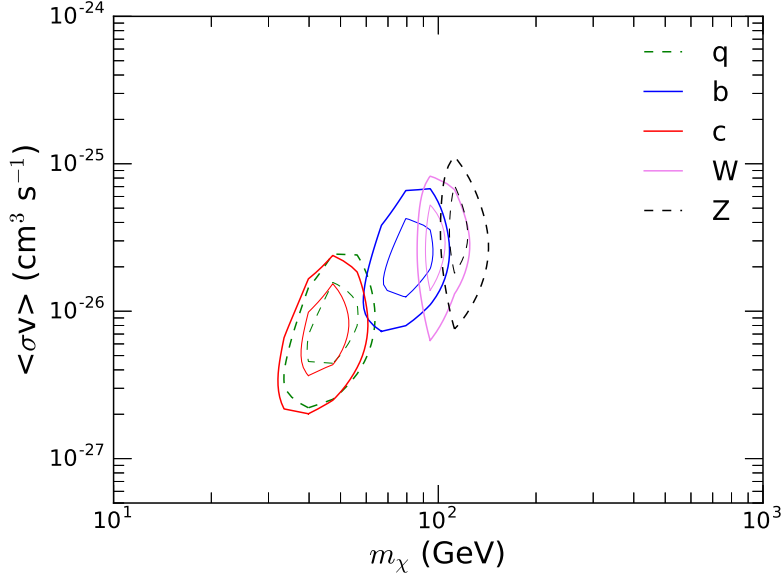}
\caption{The 68\% and 99\% credible regions on the plane of 
$(m_{\chi},\sv)$ for different DM annihilation channels obtained by fitting to the AMS-02 antiproton data. }
\label{fig:aplike}
\end{figure}

In Fig.~\ref{fig:aplike} we show the preferred regions on the $(m_{\chi},\sv)$ 
plane by the AMS-02 data given different annihilation channels. 
The DM density profile is assumed to be the Navarro-Frenk-White profile
\cite{Navarro:1996gj}. For each annihilation channel, the 68\% and 95\% 
credible contours are plotted. 
For light quarks the favored DM mass and annihilation
cross-section are $m_\chi \sim 45$ GeV and $\langle \sigma v \rangle \sim 8 \times 10^{-27}$ cm$^3$s$^{-1}$,
while heavy final states, $b$, $W$ and $Z$, need large DM masses~($\gtrsim 80$ GeV) and
cross-sections~($\gtrsim 10^{-26}$ cm$^3$s$^{-1}$).

\subsection{Likelihood of the DD data}

Given that no signal has been observed in all of the three DD experiments,
LUX~\cite{Akerib:2016vxi}, PandaX-II~\cite{Tan:2016zwf,Cui:2017nnn}
and XENON1T~\cite{Aprile:2017iyp}, we can construct the total 
likelihood function based on a combined analysis as 
\begin{eqnarray}
&&\ln \mathcal{L}_{\rm{DD}}=\sum_{i}
 \ln \mathcal{L}_{i}(m_\chi,\mathcal{R}), 
\label{Eq:Likedd}
\end{eqnarray}
where $i=\texttt{PandaX-II},~\texttt{LUX},~\texttt{XENON1T}$. As explained in 
Ref.~\cite{Liu:2017kmx}, the likelihood function of each experiment depends 
on the efficiency of the analysis cuts. Therefore, we use the event 
rate $\mathcal{R}$ (per day per kg) as the
observable instead of the cross-section.
Following Ref.~\cite{Liu:2017kmx}, we includes two types of likelihoods, 
the Poisson-Gaussian (PG) one for PandaX-II and XENON1T, 
and the Half-Gaussian (HG) one for LUX:
\begin{eqnarray}
&&\mathcal{L}_{\rm{PG}}\propto \prod_{i} 
\max_{b_i^\prime}\frac{\exp[-(s_i+b_i^\prime)] (s_i+b_i^\prime)^{o_i}}{o_i!} 
\exp\left[-\frac{(b_i^\prime-b_i)^2}{2 \delta b_i^2}\right],   \nonumber \\
&&\mathcal{L}_{\rm{HG}}\propto 
\exp\left[\frac{1}{2}\frac{s_i(\mathcal{R})}{ s_{i,{\rm{95}}}(m_\chi)/1.64}\right]^2,
\label{Eq:luxlike}
\end{eqnarray}
where $b_i$ and $\delta b_i$ are simulated background event numbers and 
their uncertainties,  $o_i$ is the observed event number reported by 
PandaX-II and XENON1T, and $s_{i,{\rm{95}}}(m_\chi)$ 
is the number of events computed from the $95\%$ confidence level 
limit in Ref.~\cite{Akerib:2016vxi}. Here, the $95\%$ confidence level 
is equivalent to $1.64 \, \sigma$ away from the central value in the 
one-dimensional Gaussian likelihood. The DD likelihood has been 
incorporated in the {\tt LikeDM} tool~\cite{Huang:2016pxg}, i.e., 
{\tt LikeDM-DD}~\cite{Liu:2017kmx}.

Finally, we can write down the posterior distribution based on the total DD likelihoods, 
 given the DM mass $m_\chi$  
\begin{equation}
\mathcal{P}_{\rm DD}(\mchi)\propto \int 
\mathcal{L}_{\rm DD}
(\mchi,\Lambda)\,
\frac{d \mathcal{R}}{d\Lambda}
\,p(\Lambda)
\,d\Lambda \, ,
\label{eq:post_DD}
\end{equation}
where we include a Jacobian $\frac{d \mathcal{R}}{d\Lambda}$ to ensure that our 
credible regions are finite. For each $\mchi$, a lower limit of
$\Lambda$ at $95\%$ CR can be inferred.

\section{Effective operators}
\label{sec:EFT}
In this section, we list the selective effective operators up to dim-6 which characterize the
interactions between the DM and  SM particles.
We here confine ourselves to 
fermion DM, denoted by $\chi$. 
The dim-5 operators are~\cite{Matsumoto:2014rxa,Bishara:2016hek}  
\begin{align}
&\mof_1 = \frac{1}{\Lambda}(\bar{\chi} \chi) (H^{\dag} H) \;\; , \;\;  \mof_2 = \frac{1}{\Lambda} (\bar{\chi} i \ga_5\chi) (H^{\dag} H)  \;\; , \nn \\
&\mof_3 = \frac{e}{8\pi^2 \Lambda}(\bar{\chi} \sig^{\mu\nu} \chi) F_{\mu\nu} \;\; , \;\; 
\mof_4 =   \frac{e}{8\pi^2 \Lambda} (\bar{\chi} i \sig^{\mu\nu} \ga_5 \chi) F_{\mu\nu} \;\; ,
\label{eq:dim-5}
\end{align}
where $H$ is the SM Higgs doublet, $F^{\mu\nu}$ is the electromagnetic 
field strength tensor and $\sigma^{\mu\nu}= i [\gamma^\mu, \gamma^\nu]/2$.
The factor of $8 \pi^2$ in the denominator is included for $\mof_3$ and $\mof_4$ as they are usually loop-induced.
The dim-6 operators are
\begin{align}
&\mos_1 = \frac{1}{\Lambda^2}(\bar{\chi} \ga_\mu \chi) (H^{\dag} i D^\mu H) \;\; , \;\;  
\mos_2 = \frac{1}{\Lambda^2} (\bar{\chi} \ga_\mu \ga_5 \chi) (H^{\dag} i D^\mu H)  \;\; ,  \nn \\
&\mos_3 = \frac{1}{\Lambda^2} (\bar{\chi} \ga_\mu \chi) (\bar{q} \ga^\mu q) \;\; , \;\;  
\mos_4 = \frac{1}{\Lambda^2} (\bar{\chi} \ga_\mu \ga_5 \chi) (\bar{q} \ga^\mu q) \;\; ,   \nn \\
&\mos_5 = \frac{1}{\Lambda^2} (\bar{\chi} \ga_\mu \chi) (\bar{q} \ga^\mu \ga_5 q) \;\; , \;\;  
\mos_6 = \frac{1}{\Lambda^2} (\bar{\chi} \ga_\mu \ga_5 \chi) (\bar{q} \ga^\mu \ga_5 q) \;\; ,
\label{eq:dim-6}
\end{align}
where $q$ explicitly refers to the SM quarks in light of the anti-proton excess,
and leptonic final states will not be considered here. 

We do not include the scalar-type operators, such as $\bar{\chi}\chi(\bar{q}q)$,
which can be generated from $\mof_1$ by integrating out the Higgs field below the 
electroweak scale. On the other hand,  tensor operators such as 
$(\bar{\chi}\sig^{\mu\nu}\chi)(\bar{q}\sig_{\mu\nu}q)$ arise from the 
$SU(2)_L$-invariant dim-7 operator, 
$(\bar{\chi}\sig^{\mu\nu}\chi)(\bar{Q}_L\sig_{\mu\nu} u_R) H$ and therefore 
will be neglected in our analysis. The $\chi$ particle can be either 
Dirac or Majorana. For Majorana $\chi$, the operators $\mof_3$, $\mof_4$, 
$\mos_1$, $\mos_3$ and $\mos_5$ will vanish. 
Interestingly, operator $\mos_5$ with Dirac DM could explain the Galactic center gamma-ray excess as well~\cite{Alves:2014yha}.

As pointed out in Refs.~\cite{Kurylov:2003ra, Giuliani:2005my,Chang:2010yk,
Kang:2010mh,Feng:2011vu,Gao:2011bq} and recently demonstrated in Ref.~\cite{Liu:2017kmx}, a certain ratio 
of DM-neutron coupling to that of DM-proton ($c_n/c_p$) can significantly suppress the DM-nuclei scattering rate  
owing to the destructive interference between the proton and neutron contributions, so-called
maximum iso-spin violation~(ISV).
In case of  $c_n/c_p=1$, one has iso-spin conservation~(ISC). 
Although a precise value of $c_n/c_p$ for maximum ISV depends on the experimental setup, 
an approximate ratio can still be estimated according to the published efficiency 
from PandaX-II, LUX, and XENON1T as shown in Ref.~\cite{Liu:2017kmx}.
Because of the maximum cancellation, the weakest DD
limit in maximum ISV can be regarded as a conservative bound.

To facilitate analysis, the ratio $c_n/c_p$ is translated into the coupling ratio of DM-up type quark to
DM-down type quark, $c_u/c_d$. 
We note that different values of $c_u/c_d$ 
will also slightly shift AMS02 favored regions in the plane of $m_\chi$ and $\Lambda$.
For computation of the DM event rate, we employ the \texttt{DirectDM} code~\cite{Bishara:2017nnn} 
to RG evolve the effective couplings from the electroweak scale
 down to the scale slightly higher than the QCD scale, and then match to the DM-nucleon coupling 
by using the chiral perturbation theory. 
That can be done via the existing tools in order:  
\texttt{DirectDM}~\cite{Bishara:2017nnn} $\to$ \texttt{DMFormFactor}~\cite{Anand:2013yka,Fitzpatrick:2012ix} 
$\to$ \texttt{LikeDM-DD}~\cite{Liu:2017kmx}.

\begin{figure}[!htb]
\includegraphics[width=0.45\textwidth]{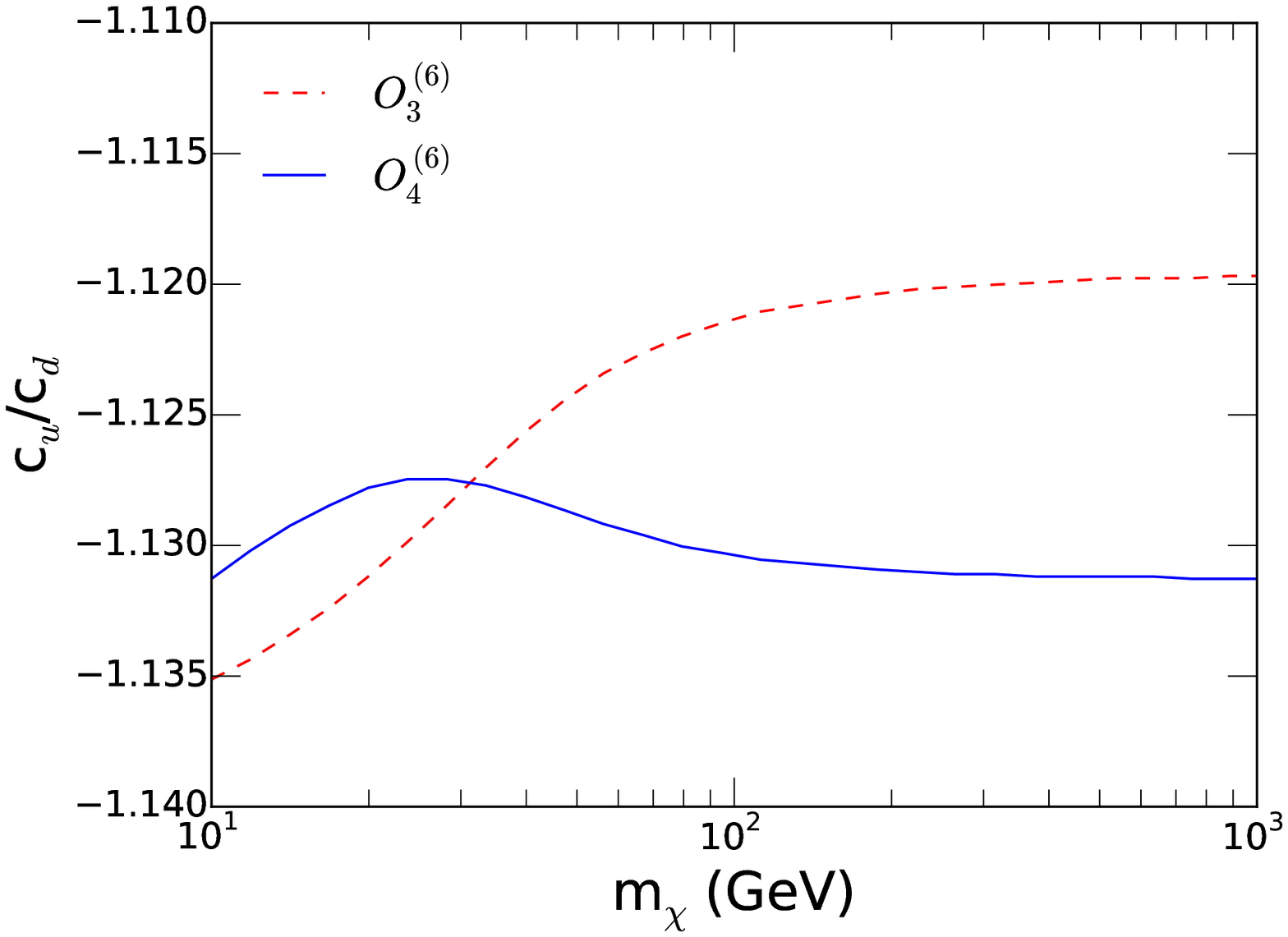}
\includegraphics[width=0.45\textwidth]{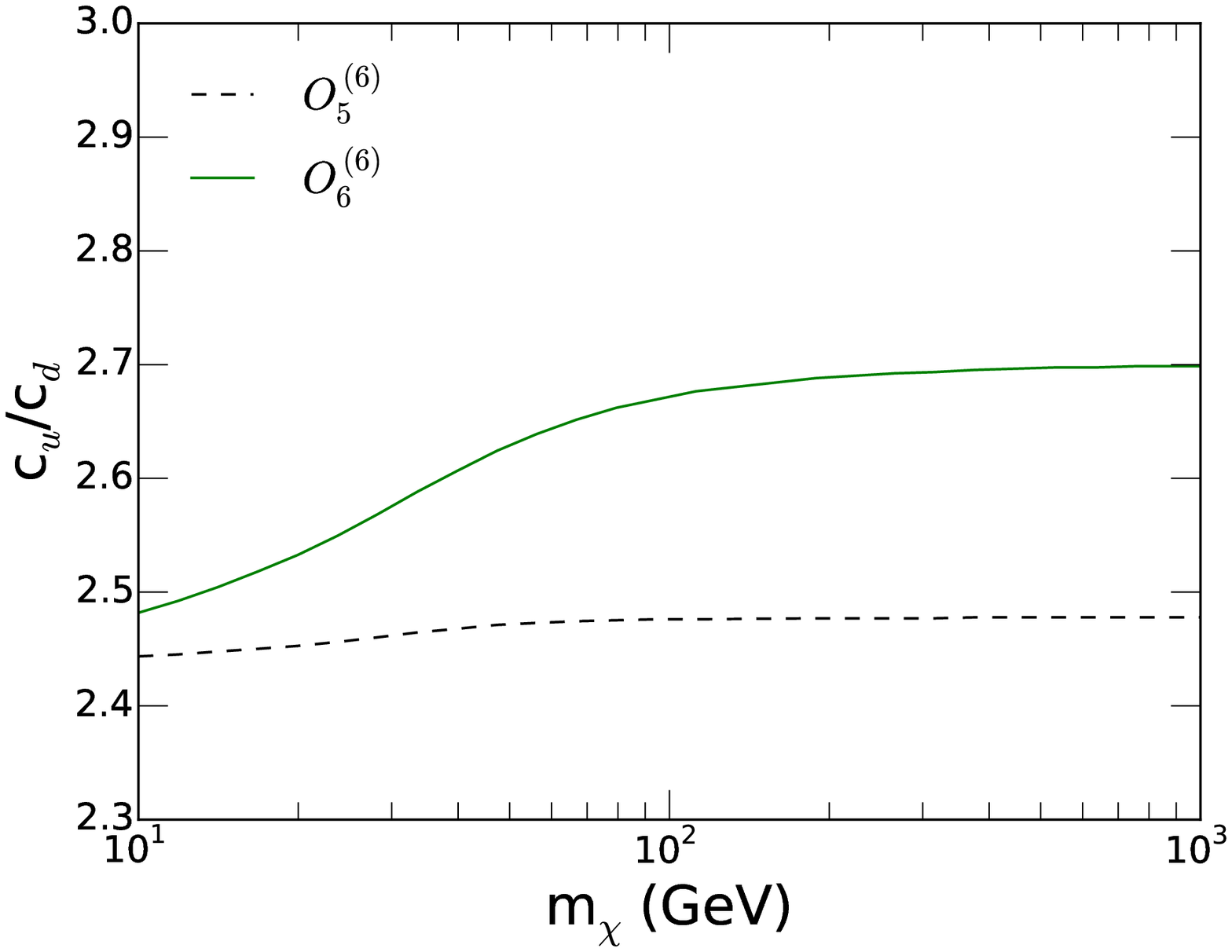}
\caption{The ratio $c_u/c_d$ for obtaining maximal ISV as a function of the DM mass
for operators $\mos_{3,4}$~(left panel) and $\mos_{5,6}$~(right panel).  
\label{fig:ISV}}
\end{figure}

For simplicity, we assume an universal $c_u$ for all the up-type quarks and 
$c_d$ for all the down-type quarks. 
At quark level, for the dim-5 operators the ratio $c_u/c_d$ is fixed by either the (quark-)Yukawa
or gauge couplings in light of the Higgs and gauge bosons involved in the operators. 
To be more precise, $c_u/c_d$ for operators $\mof_1$ and $\mof_2$  is 
the up-type Yukawa coupling divided by the down-type one, while 
for operators $\mof_3$ and $\mof_4$, $c_u/c_d= -2$.

Except for $\mos_1$ and $\mos_2$, whose couplings are simply 
 the SM Yukawa or gauge couplings, the dim-6 operators
the ratio $c_u/c_d$ is unconstrained.        
In Fig.~\ref{fig:ISV}, we display the required $c_u/c_d$ 
to achieve  maximum ISV. The ratio depends on the DM mass 
but remains constant for large DM masses.

Finally, we should point out that EFT is only valid if the mediator is much heavier than DM, i.e., the cut-off scale should be larger than twice the DM mass: $\Lambda > 2\, m_\chi$, assuming the relevant couplings are of $\mathcal{O}(1)$.
For those operators involving $H$, $\mof_{1,2}$ and $\mos_{1,2}$,
there exists another scale -- the Higgs
vacuum exception value (VEV) -- as one of the Higgs fields
can be replaced by the VEV while the rest one is integrated out  below the electroweak scale. As a result, for these operators we require 
\begin{equation}
\Lambda > \max\left[2 m_\chi, 174\, \text{GeV} \right],
\label{eq:lambdalimit}
\end{equation}
to avoid an unphysical enhancement from large values of $\langle H \rangle/\Lambda$.
On the other hand, as long as $\Lambda$ is larger than the scale of DM-nuclei momentum exchange~($\sim 100$ MeV),
the derived DD bound based on EFT is reliable, provided that the RG evolution and the matching to the nuclear theory are properly
included.

\section{Results}
\label{sec:result}

\begin{table}[t]
\begin{tabular}{|c|c|c|c|}
\hline
\hline
operator & Annihilation~(ID) & DM-nuclei scattering~(DD) & Dirac/Majorana\\
\hline\hline
$\mof_1 = \frac{1}{\Lambda}(\bar{\chi} \chi) (H^{\dag} H)$ & $p$-wave  & $\mathcal{Q}_1$ & 
YES/YES\\
$\mof_2 =  \frac{1}{\Lambda}\bar{\chi} i \ga_5\chi H^{\dag} H$ & $s$-wave  &  $\mathcal{Q}_{11}$ &
YES/YES\\
$\mof_3 = \frac{e}{8\pi^2\Lambda}(\bar{\chi} \sig^{\mu\nu} \chi) F_{\mu\nu}$ & $s$-wave  & $\mathcal{Q}_{1,4,5,6}$ 
& YES/NO\\
$\mof_4 = \frac{e}{8\pi^2\Lambda} (\bar{\chi}i\sig^{\mu\nu}\ga_5 \chi) F_{\mu\nu}$ & $p$-wave & $\mathcal{Q}_{11}$ 
& YES/NO\\
\hline
\hline
$\mos_1 =\frac{1}{\Lambda^2}(\bar{\chi} \ga_\mu \chi) (H^{\dag} i D^\mu H)$ & $s$-wave  & $\mathcal{Q}_{1,7,9}$  
& YES/NO\\
$\mos_2 =\frac{1}{\Lambda^2}(\bar{\chi} \ga_\mu \ga_5 \chi) (H^{\dag} i D^\mu H)$ & $s$-wave & 
$\mathcal{Q}_{4,6,8,9}$ & YES/YES\\
$\mos_3 =\frac{1}{\Lambda^2}(\bar{\chi} \ga_\mu \chi) (\bar{q} \ga^\mu q)$ & $s$-wave  & $\mathcal{Q}_{1}$ &  
YES/NO\\
$\mos_4 =\frac{1}{\Lambda^2}(\bar{\chi} \ga_\mu \ga_5 \chi) (\bar{q} \ga^\mu q)$ & $p$-wave & $\mathcal{Q}_{8,9}$&
YES/YES\\
$\mos_5 =\frac{1}{\Lambda^2}(\bar{\chi} \ga_\mu \chi) (\bar{q} \ga^\mu \ga_5 q)$ & $s$-wave & $\mathcal{Q}_{7,9}$& 
YES/NO\\
$\mos_6 =\frac{1}{\Lambda^2}(\bar{\chi}\ga_\mu \ga_5\chi)(\bar{q} \ga^\mu \ga_5 q)$ & $s$-wave~($m^2_q$) & 
$\mathcal{Q}_{4,6}$ &YES/YES\\
\hline
\hline
\end{tabular}
\caption{
The summary of the selective operators' properties in terms of ID and DD.  For $\mos_6$,
the annihilation cross-section is proportional to the final-state fermion mass as indicated by $m^2_q$, and thus suppressed for light
quarks. 
The operators $\mathcal{Q}_i$ is DM non-relativistic operators for DD, 
defined in Appendix~\ref{sec:appC}.
}
\label{table:summary}
\end{table} 

In this section, with the help of a modified version of the
\texttt{LikeDM} package~\cite{Liu:2017kmx} we investigate if any of the  operators in Eqs.~\eqref{eq:dim-5} and \eqref{eq:dim-6}
 can simultaneously explain the AMS02 antiproton excess and satisfy 
the combined bounds from PandaX-II, LUX, and XENON1T.
Our results are presented in the plane of the DM mass $m_\chi$ and the cut-off scale $\Lambda$. 
We utilize \texttt{FeynRules}~\cite{Alloul:2013bka} to generate the interaction vertices which are then
imported to \texttt{MicrOMEGAs}~\cite{Barducci:2016pcb} for  annihilation cross-section computations,
 except for two dim-5 operators, $\mof_3$ and $\mof_4$ whose cross-sections are manually computed and collected in 
Appendix~\ref{sec:appB}. Furthermore, for the dim-6 operators
we display both ISC and ISV DD constraints to demonstrate  
how much the limits can be mitigated by ISV.  

In Table~\ref{table:summary}, we summarize the properties of
all the effective operators in terms of direct and indirect detection. 
The operators $\mathcal{Q}_i$ are the non-relativistic  DM-nucleus  operators 
defined in Appendix~\ref{sec:appC}.
Particularly, 
 $\mathcal{Q}_{1}$ and $\mathcal{Q}_{4}$ are the SI and spin-dependent~(SD) operators respectively,
and are the only operators without velocity or momentum dependence~(i.e, without suppression).

\subsubsection*{Dimension-five operators}

\begin{figure}[!htb]
\includegraphics[width=0.45\textwidth]{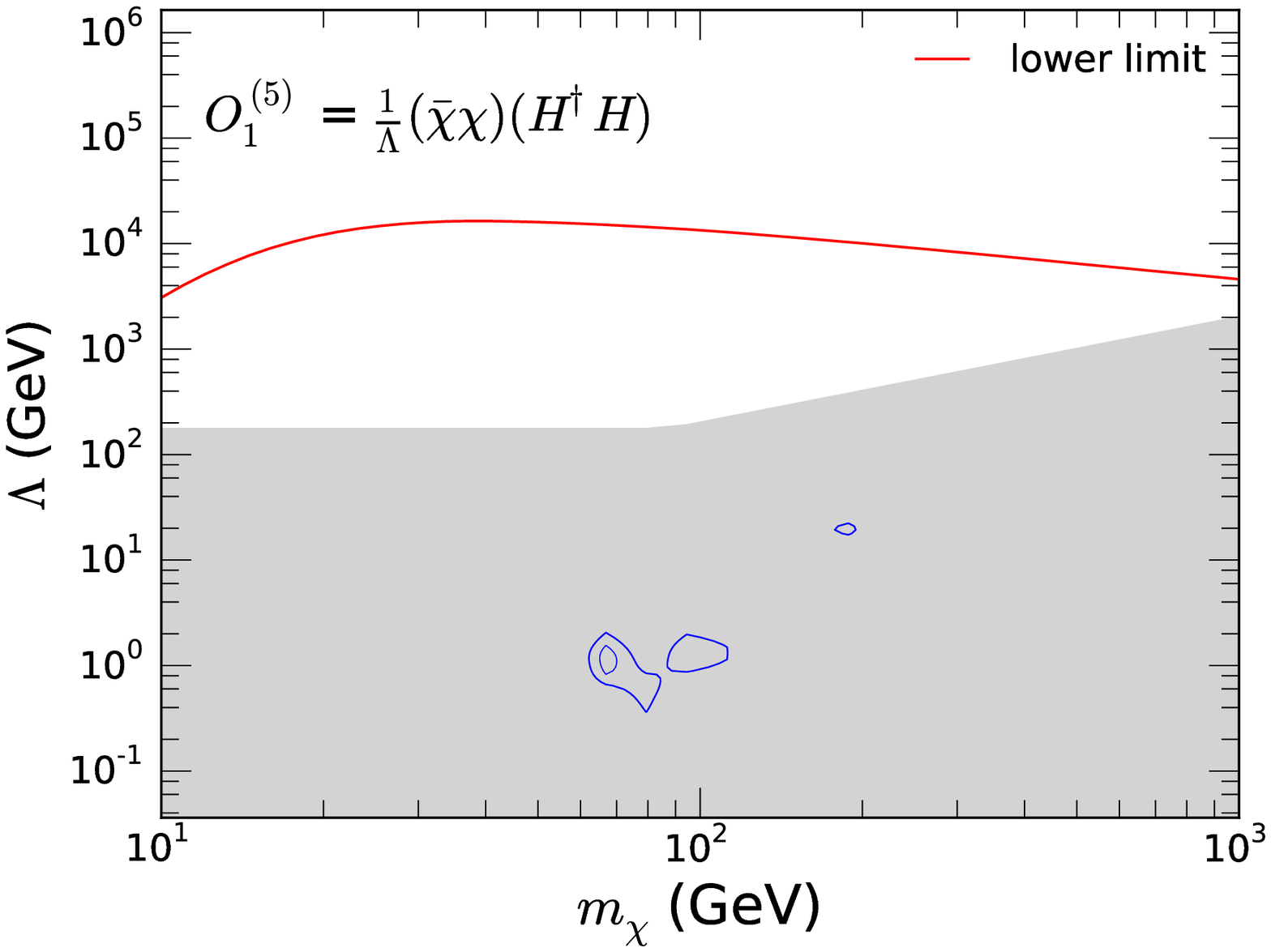}
\includegraphics[width=0.45\textwidth]{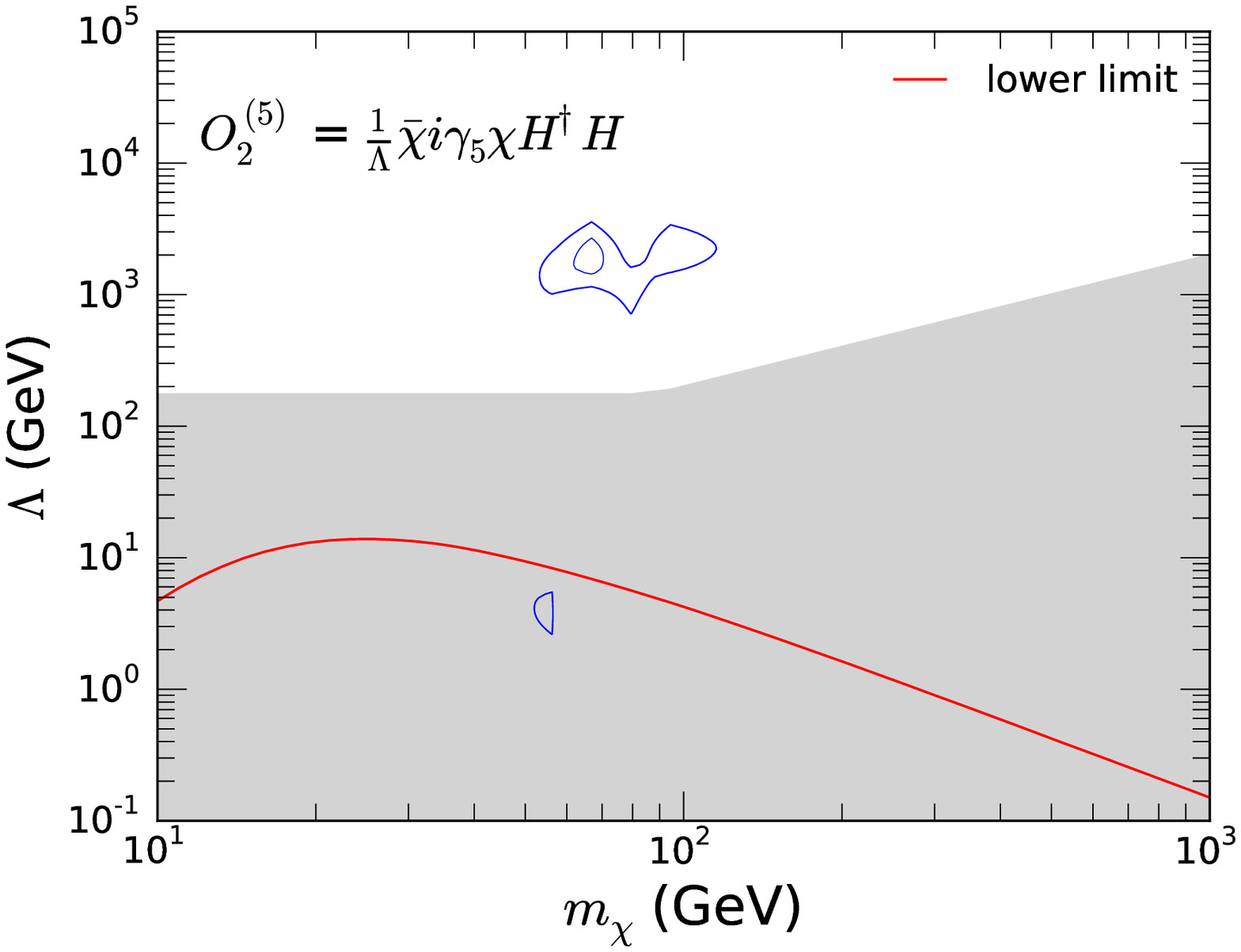}\\
\includegraphics[width=0.45\textwidth]{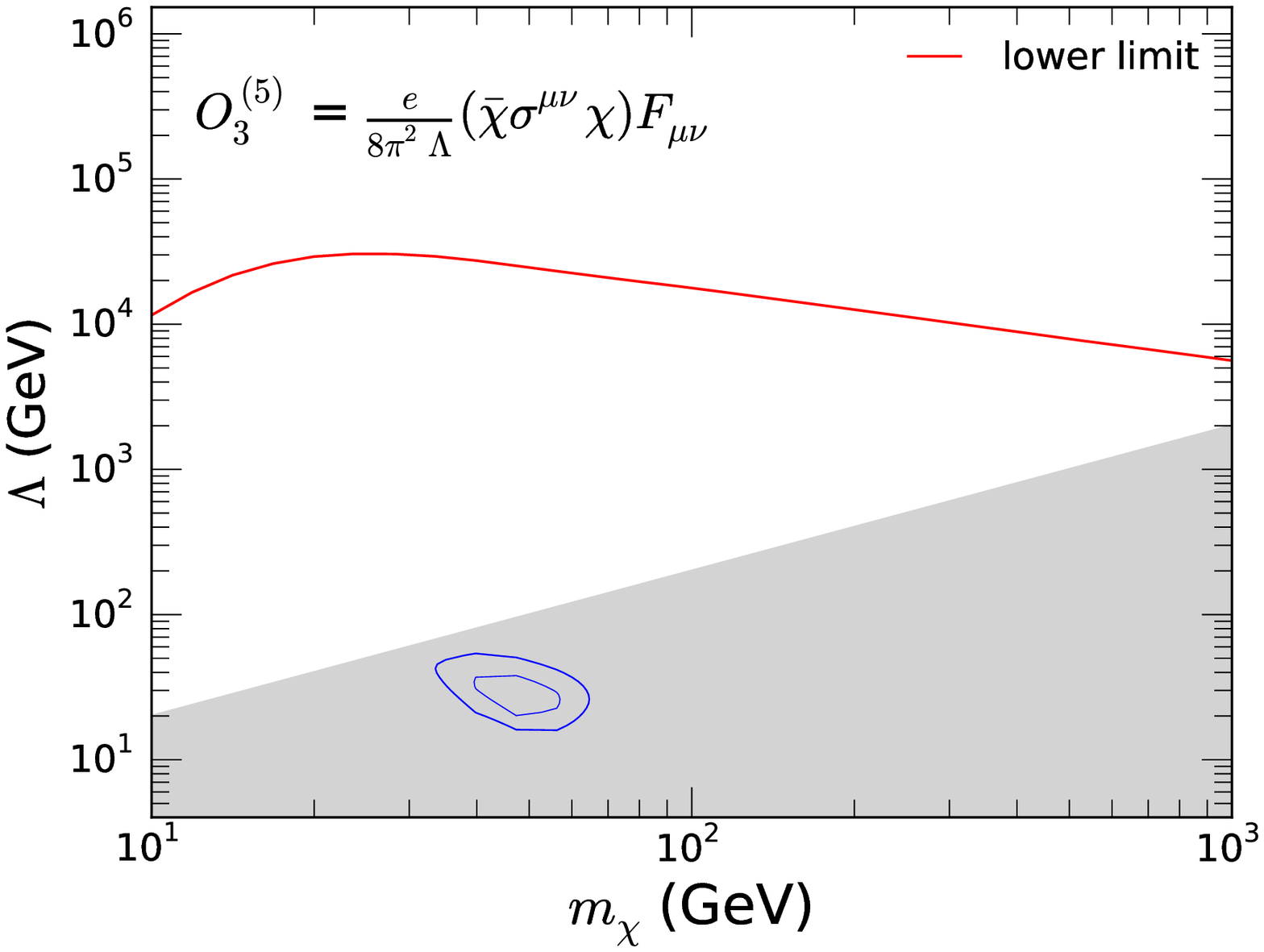}
\includegraphics[width=0.45\textwidth]{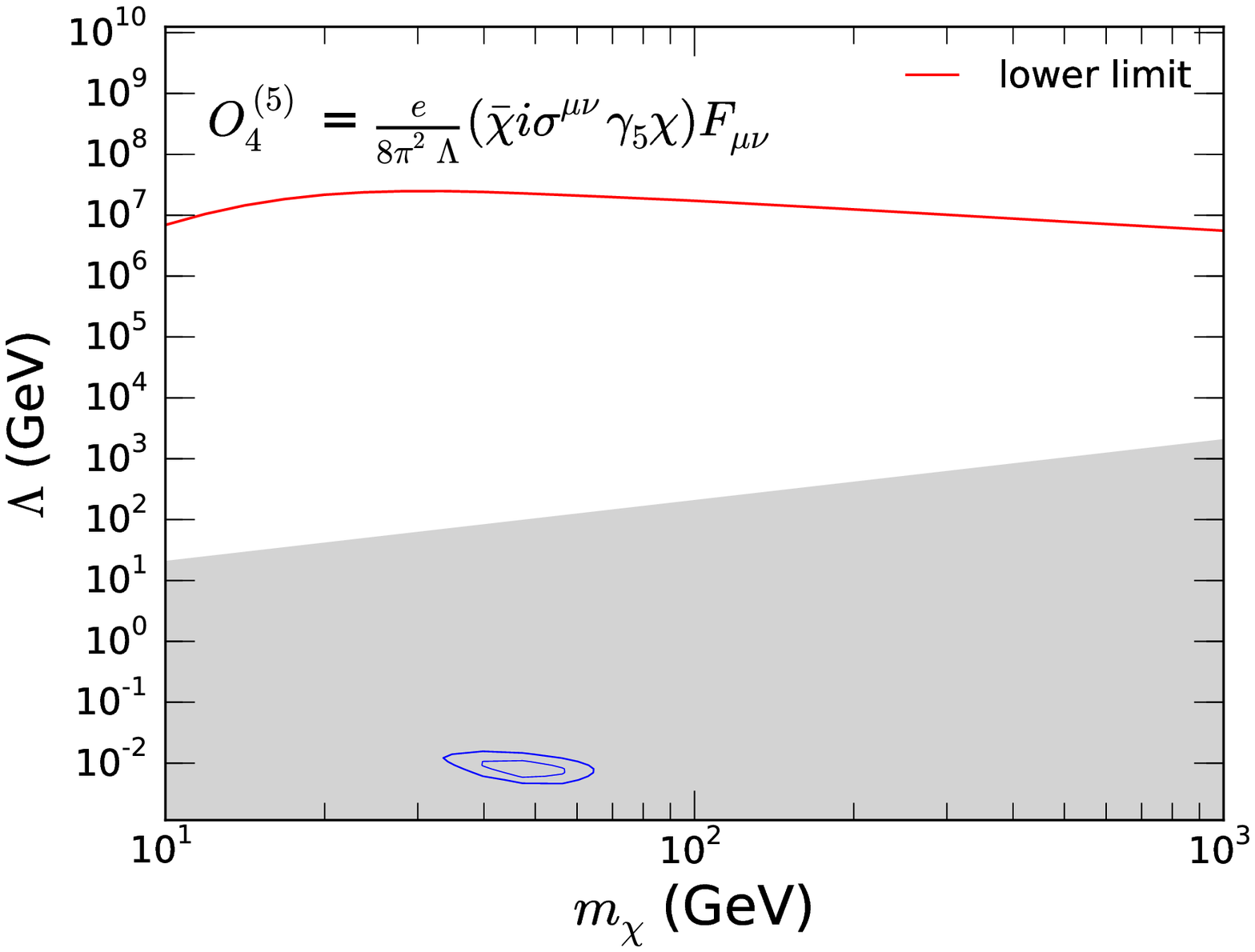}\\
\caption{
The result for $\mof_1$ and $\mof_2$ (upper-left and upper-right) and
$\mof_3$ and $\mof_4$ (bottom-left and bottom-right). 
The red curves denote the lower limit on $\Lambda$
at $95\%$ CR derived from the combined DD measurements, 
while the blue contours represent the $68\%$ (inner) and $95\%$ (outer) 
favored region by the AMS02 antiproton data. 
\label{fig:O5}}
\end{figure}

In Fig.~\ref{fig:O5}, the two upper panels correspond to $\mof_1$ and $\mof_2$ while  
the lower ones are $\mof_3$ and $\mof_4$. The red curve corresponds to  
the lower limit on $\Lambda$ at $95\%$ CR from the combined DD measurements; 
namely the region below the red curve is excluded.
The blue contours stand for the $68\%$ (inner) and $95\%$ (outer) CR preferred by the AMS02 antiproton data. 
The EFT approach for DM annihilation cross-section computation becomes unreliable within the grey shaded
area as mentioned above in Eq.~\eqref{eq:lambdalimit}.
Moreover, in Table~\ref{table:O5} we list the annihilation branching fractions for few benchmark points that are located in the best-fit regions to highlight dominant contributions. 

Apart from operator $\mof_2$,
the AMS02 favored regions lie within the grey area
as the corresponding annihilation cross-sections are either suppressed by the small DM velocity~($\mof_1$ and $\mof_4$) and/or
by the loop suppression~($\mof_3$ and $\mof_4$)  which needs a small value of $\Lambda$ to counterbalance. 
In other words, the EFT results are not accurate in these cases and effects of mediators which connect the DM and
SM sectors, such as the resonance enhancement, have to be taken into account. 
By contrast, the lower bounds on $\Lambda$ derived from the combined DD data are above the DM-nuclei
momentum-transfer scale~($\lesssim$ 100 MeV), implying EFT computation of DM-nuclei scattering is legitimate.

\begin{table} 
\scriptsize
\centering
\begin{tabular}{| l | l | l |  l | l | l | l|l|l|}
\hline
\hline
& $\mof_1$-low & $\mof_1$-mid & $\mof_1$-high &$\mof_2$-low & $\mof_2$-mid & $\mof_2$-high & $\mof_3$ & $\mof_4$ \\
\hline
\multicolumn{9}{|c|}{Mass Parameters}\\
\hline
$\mchi$ (\gev) &  70  & 100 & 190  & 55 &	65  & 100 & 50 & 50 \\
$\Lambda$ (\gev) &  1 & 1 & 20   & 3.5  & $1.8\times 10^3$ & $2\times 10^3$ & 25 & $8\times10^{-3}$\\
\hline
\multicolumn{9}{|c|}{Annihilation Branching Ratio}\\
\hline

$\bar{\chi}\chi\to b\bar{b}$
&  $59.5\%$     
&  $<1 \%$ 
&  $0 \%$ 
&  $86.8\%$  
&  $70.4\%$
&  $<1 \%$ 
&  $2.63\%$
&  $2.63\%$
\\

$\bar{\chi}\chi\to c\bar{c}$
&  $4.37\%$     
&  $<1 \%$ 
&  $0 \%$ 
&  $6.4\%$
&  $5.18\%$  
&  $<1 \%$
&  $ 10.5\%$
&  $ 10.5\%$
\\

$\bar{\chi}\chi\to \tau^+\tau^-$
&  $2.85\%$     
&  $<1 \%$
&  $0 \%$ 
&  $4.18\%$
&  $3.38\%$ 
&  $<1 \%$
&  $23.68\%$
&  $23.68\%$
\\

$\bar{\chi}\chi\to ZZ$
&  $3.78\%\,^*$     
&  $25.8 \%$ 
&  $<1 \%$
&  $0\%$
&  $2.3\%\,^*$ 
&  $25.8\%$
&  $0\%$
&  $0\%$
\\

$\bar{\chi}\chi\to W^+W^-$
&  $29.4\%\,^*$     
&  $73.5 \%$ 
&  $<1 \%$
&  $2.54\%\,^*$
&  $18.7\%\,^*$
&  $73.5\%$
&  $15.79\%$ 
&$15.79\%$
\\

$\bar{\chi}\chi\to HH$
&  $0\%$
&  $0\%$
&  $99.7 \%$
&  $0\%$
&  $0\%$
&  $0\%$
&  $0\%$
& $0\%$
\\

\hline
\hline
\end{tabular}
\caption{\label{table:O5} 
The branching fractions of the final states of interest for dim-5 operators. 
The off-shell contribution is marked by the star symbol $*$. Note that the total branching ratios are not equal to
unity as leptonic channels also exist. 
}
\end{table}

There are more than one favored region for operators $\mof_1$ and $\mof_2$ 
due to the Higgs VEV insertion.      
The dominate channels for $\mof_1$ are $\bar{\chi}\chi \to H \to b\bar{b}$
(for $m_\chi \sim 70\gev$), 
$\bar{\chi}\chi \to H \to W^+W^-$ ($100\gev$), and $\bar{\chi}\chi \to HH$ ($190\gev$).   
We here also include the sizable off-shell contributions from the $WW^*$ and $ZZ^*$ final states
when DM is lighter than $W$ or $Z$.
The operator $\mof_2$ features the $s$-wave annihilation cross-section and thus larger $\Lambda$ 
than  the $p$-wave suppressed $\mof_1$. 
Besides, there are only two dominant channels, $b\bar{b}$ and $W^+W^-$ for the best-fit regions. 
The tiny best-fit region with $\Lambda \sim 3.5$ GeV is due to the unphysical enhancement from
$\langle H \rangle / \Lambda \gg 1$ as mentioned above.

The operators $\mof_3$ and $\mof_4$~(lower panels of  Fig.~\ref{fig:O5}) 
correspond to the DM magnetic and electric dipole moment interactions respectively.
The favored region of $\mof_4$ has much lower values of $\Lambda$ than that of $\mof_3$ as
$\mof_4$ is $p$-wave suppressed.
The dominant annihilation channels for $\mof_3$ and $\mof_4$, nonetheless, are identical, leading to  
the same DM mass range. In addition, although $\mof_4$ has a velocity-suppressed SI interaction, because of
the RG running of the Wilson coefficient together with the matching, the SI interaction of $\mof_4$
is enhanced at low energies and actually becomes stronger than that of $\mof_3$
which has an unsuppressed SI interaction~\cite{Liu:2017kmx}.
In other words, $\mof_4$ has a more stringent constraint on $\Lambda$ than $\mof_3$ as can be seen from Fig.~\ref{fig:O5}.

All in all, except for $\mof_2$
all the best-fit regions fall into the regime where the EFT approach
is no long valid. In general, in the absence of any cross-section boost mechanism such as
resonance enhancement, the annihilation
cross-section based on EFT should not be very different from that of the corresponding
underlying theory. In other words, due to the fact $\mof_{1,3,4}$ have the best-fit regions far below the DD bound,
 they will most likely be ruled out by direct searches unless, for example,
 an enormous resonance enhancement is involved, implying severe fine-tuning.

\subsubsection*{Dimension-six operators}

\begin{figure}[!htb]
\includegraphics[width=0.45\textwidth]{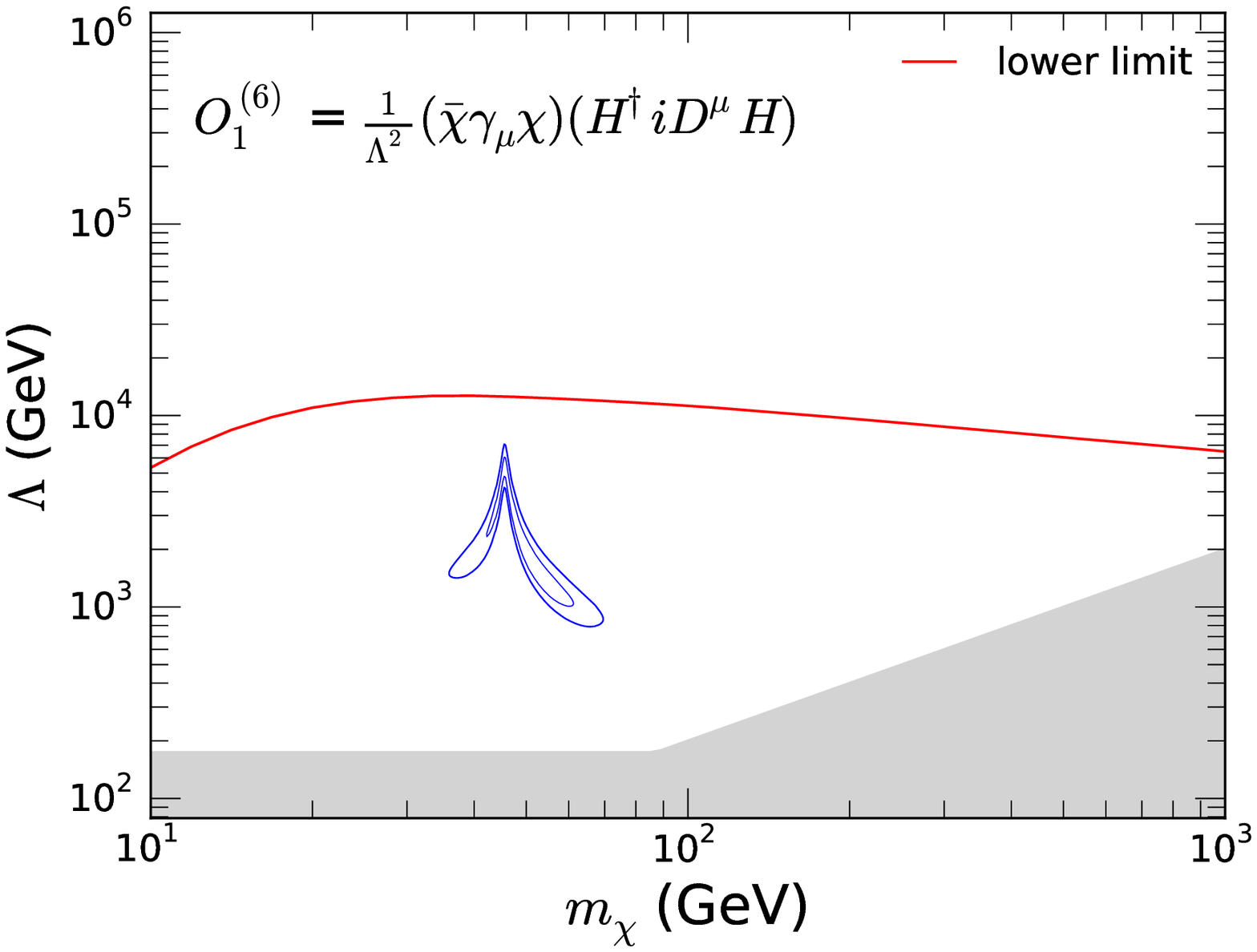}
\includegraphics[width=0.45\textwidth]{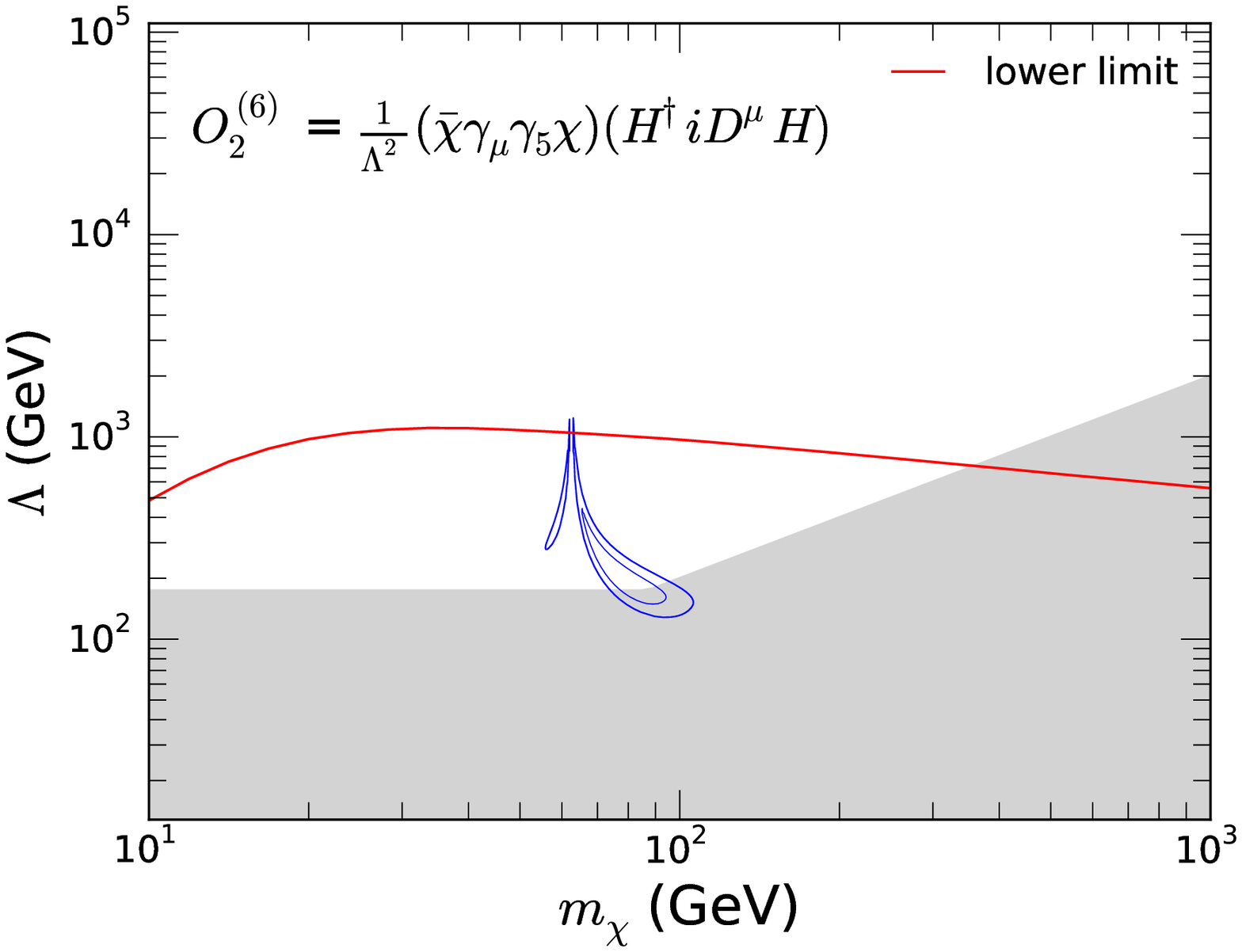}\\
\includegraphics[width=0.45\textwidth]{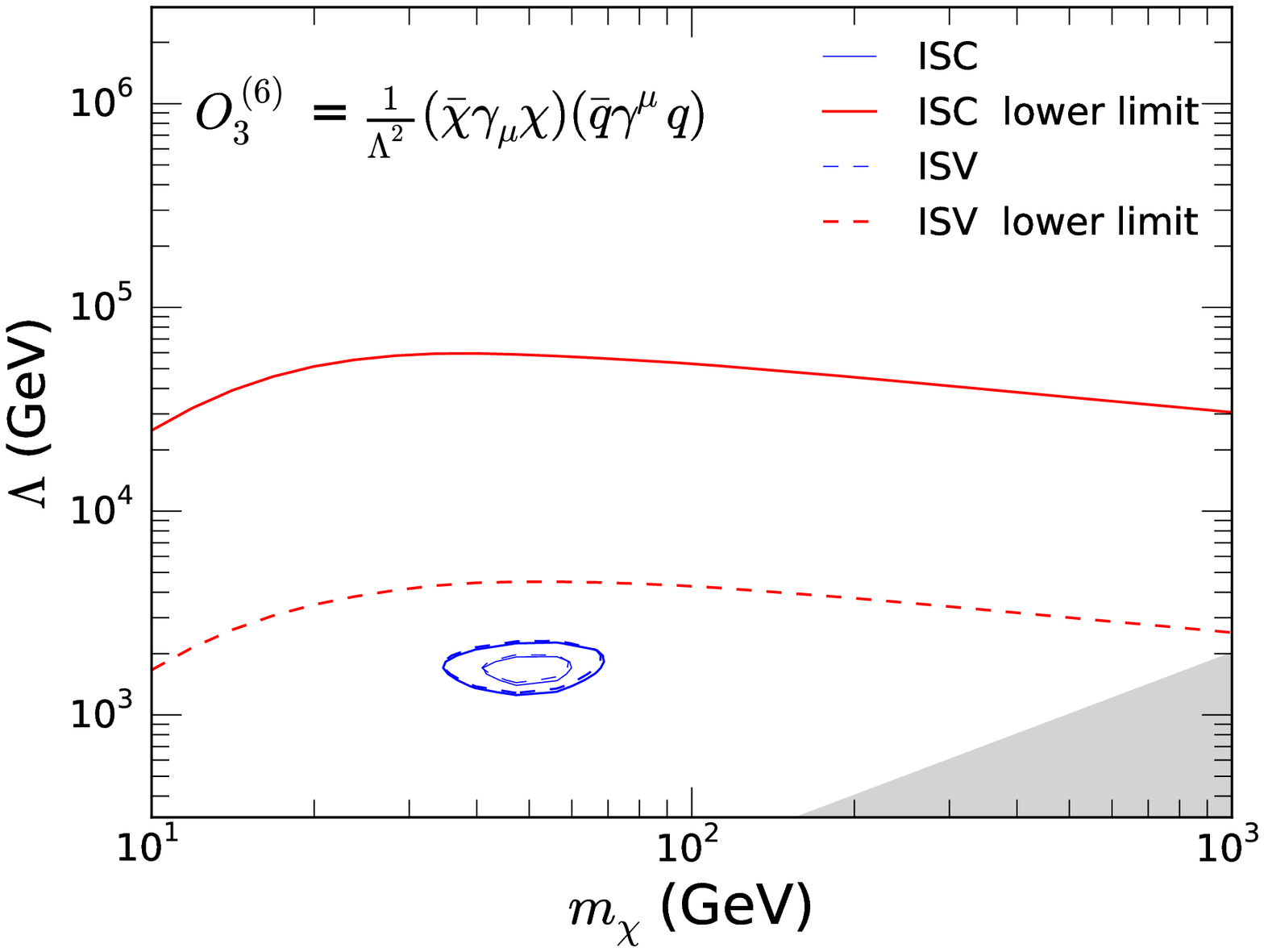}
\includegraphics[width=0.45\textwidth]{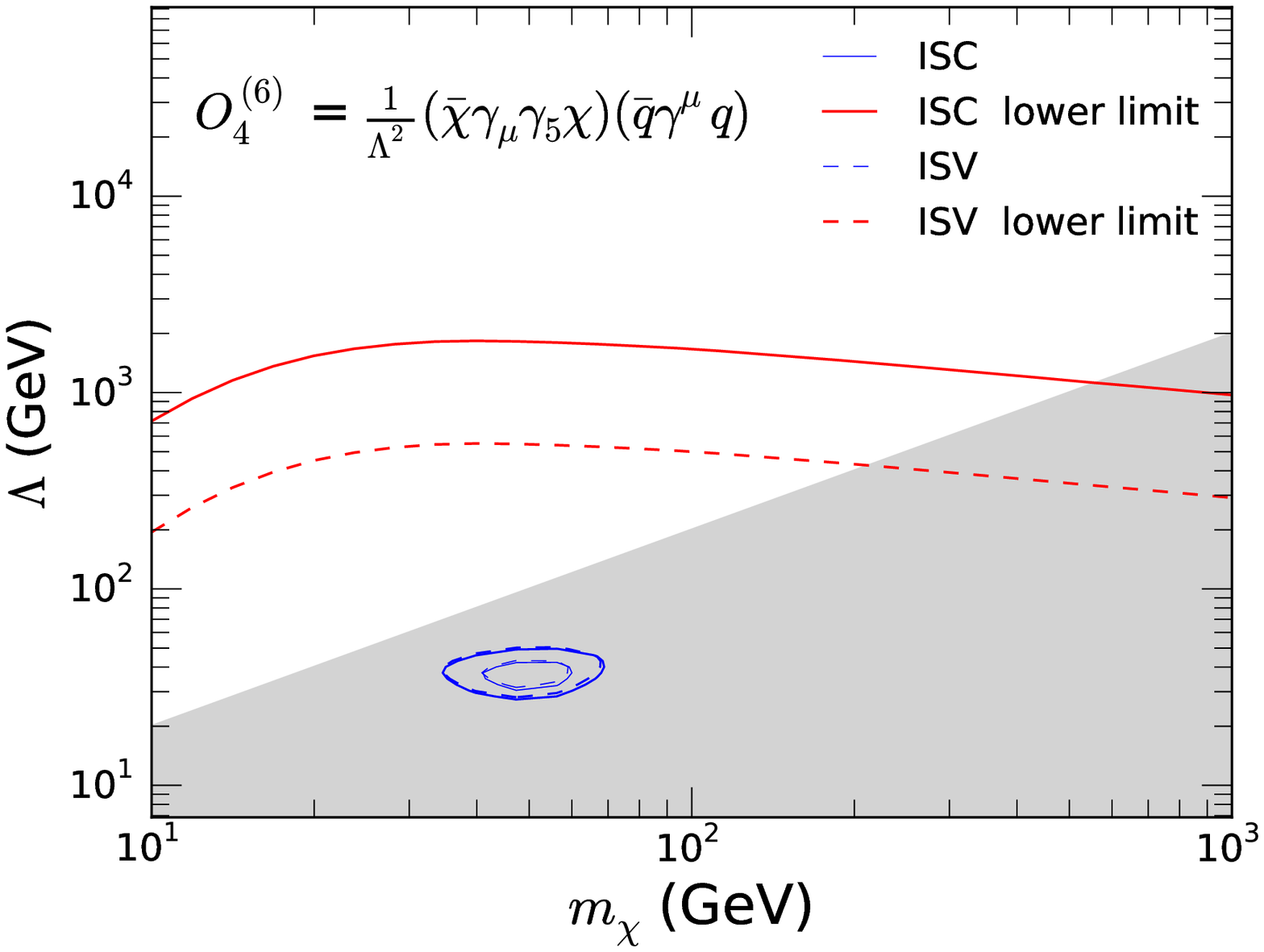}\\
\includegraphics[width=0.45\textwidth]{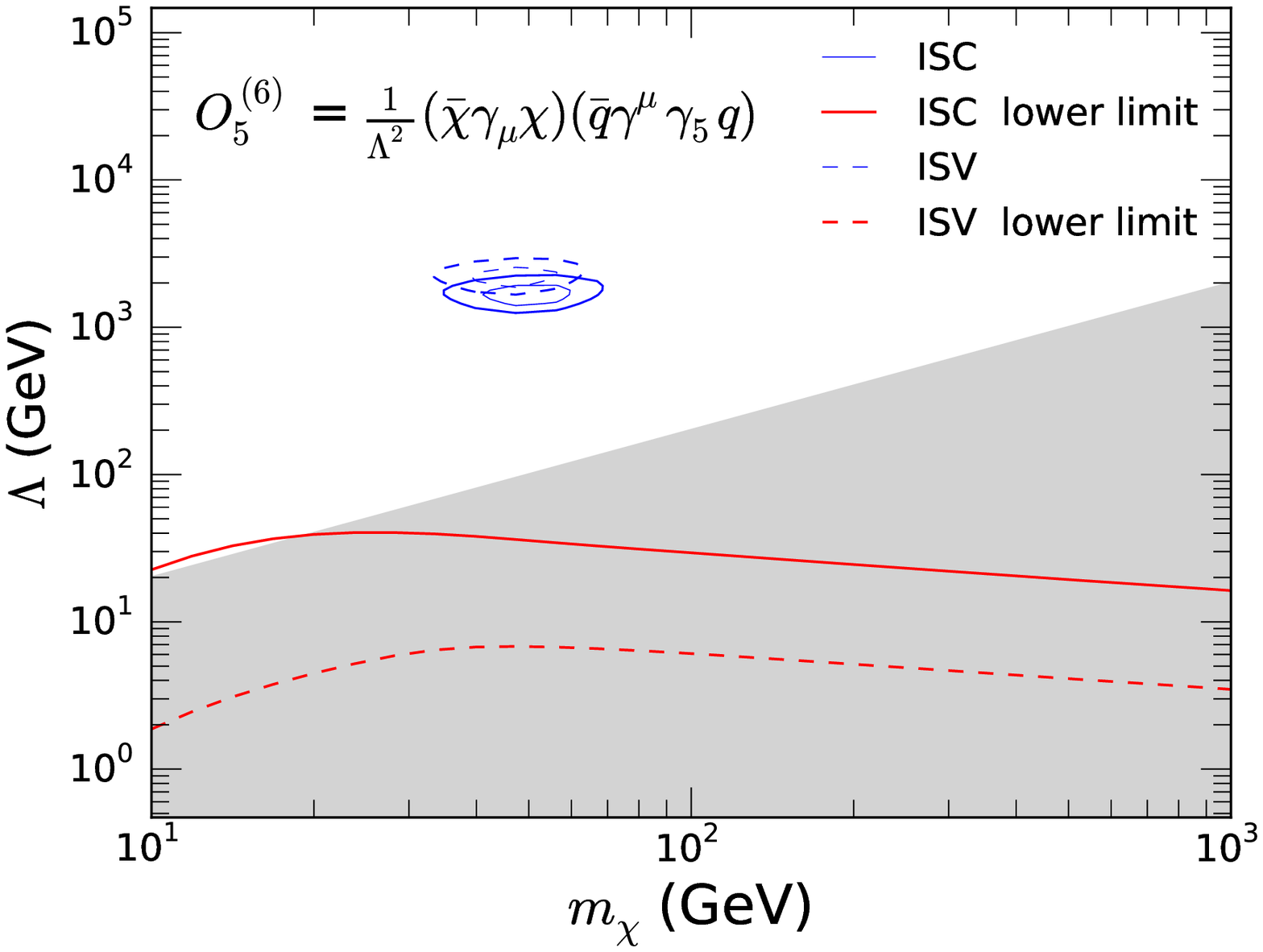}
\includegraphics[width=0.45\textwidth]{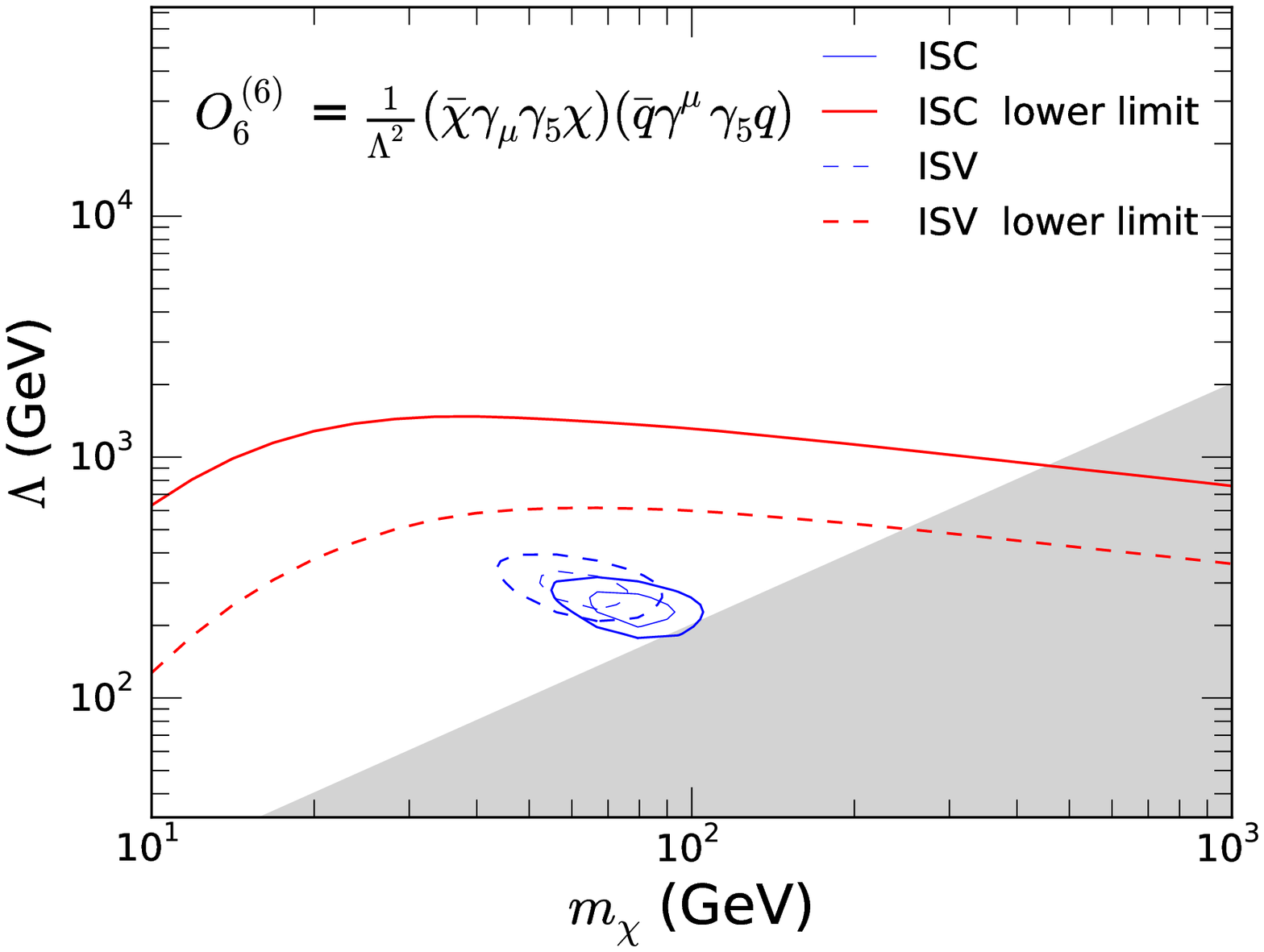}\\
\caption{
The DM direction detection bounds (red) and AMS02 antiproton excess (blue)
based on the ISC (solid) and ISV (dashed) scenarios. 
\label{fig:O6}}
\end{figure}

In Fig.~\ref{fig:O6}, we present the DD bounds for the ISC (solid red) and maximum ISV (dashed red) scenarios
and also the best-fit regions to the AMS02 antiproton excess (blue).
Due to the maximal cancellation, the limits on $\Lambda$
can be lowered by roughly one order of magnitude while the favored regions are not very sensitive to 
values of  the ratio $c_u/c_d$ -- in some cases ISV features slightly larger $\Lambda$.
Note that similar to the dim-5 operators, DM-nuclei scattering cross-section computations
based on EFT are valid since the momentum transfer is much smaller than $\Lambda$.
As in the dim-5 case, the branching ratios of the final states are listed in Table~\ref{table:O6} for few benchmark points within the
best-fit regions and only ISC is shown.

\begin{table} 
\scriptsize
\centering
\begin{tabular}{|l|l|l|l|l|l|l|l|l|l|}
\hline
\hline
& $\mos_1$-low & $\mos_1$-mid &$\mos_1$-high & $\mos_2$-low & $\mos_2$-high & 
$\mos_3$ & $\mos_4$ & $\mos_5$ & $\mos_6$ \\
\hline
\multicolumn{10}{|c|}{Mass Parameters}\\
\hline
$\mchi$ (\gev) &  40.0 & 50.0 & 60.0 &	65.0  & 80.0 & 50.0 & 50.0 & 50.0 & 80.0\\
$\Lambda$ (\gev) &  $1.8\times 10^{3}$  & $3\times 10^{3}$ & $10^3$ & $350.0$ 
& $190.0$ & $1.6\times 10^{3}$ & $35.0$& $1.6\times10^3$ &$230.0$\\
\hline
\multicolumn{10}{|c|}{Annihilation Branching Ratio}\\
\hline

$\bar{\chi}\chi\to b\bar{b}$
&  $15.0\%$     
&  $15.1\%$ 
&  $15.1\% $ 
&  $88.4\%$  
&  $<1\%$
&  $20\%$
&  $20\%$ 
&$19.8\%$
&$93.1\%$
\\

$\bar{\chi}\chi\to c\bar{c}$
&  $11.8\%$     
&  $11.8\% $ 
&  $11.8\% $
&  $6.57\% $  
&  $0\%$
&  $20\%$
&  $20\%$ &$20\%$&$6.82\%$
\\

$\bar{\chi}\chi\to q\bar{q}$
&  $42.2\%$     
&  $42.2\% $
&  $42.2\% $
&  $<1\%$ 
&  $0\%$
&  $60\% $
&  $60\% $ &$60.3\%$&$<1\%$
\\

$\bar{\chi}\chi\to t\bar{t}$
&  $0\%$     
&  $0\% $
&  $0\% $
&  $0\% $ 
&  $17.4\%$
&  $0\%$
&  $0\% $ &$0\%$&$0\% $
\\

$\bar{\chi}\chi\to \tau^+\tau^-$
&  $3.45\%$     
&  $3.45\% $
&  $3.45\% $
&  $4.26\%$ 
&  $0\%$
&  $0\%$
&  $0\% $ &$0\%$&$0\% $
\\

$\bar{\chi}\chi\to ZZ$
&  $0\%$     
&  $0\% $ 
&  $0\%  $
&  $0\% $ 
&  $17.2\%$
&  $0\%$
&  $0\% $ &$0\% $&$0\% $
\\

$\bar{\chi}\chi\to W^+W^-$
&  $0\%$     
&  $0\% $ 
&  $<1\% $
&  $<1\% $
&  $37.4\%$
&  $0\%$
&  $0\% $ &$0\% $&$0\% $
\\

\hline
\hline
\end{tabular}
\caption{\label{table:O6} 
The list of annihilation channels for dim-6 operators. 
Only ISC cases are presented and the channel $q\bar{q}$ represents  
those of light quarks, $u$, $d$, and $s$ combined.
}
\end{table}

For operators $\mos_1$ and $\mos_2$, DM can annihilate into SM particles via $H$~(with one Higgs VEV),
or $Z$ exchange~(two Higgs VEVs) in addition to  $\bar{\chi} \chi \to H H$. The favored region of $\mos_1$
is dominated by the $Z$ exchange as can be seen from Table~\ref{table:O6} that the branching ratios of the quarks are comparable
and the resonance enhancement~(larger $\Lambda$ for compensation) occurs around $m_\chi= m_Z/2$.
By contrast, the processes $\bar{\chi}\chi \to H \to \bar{b} b, \, W^+ W^- , \, ZZ$ are important for $\mos_2$, and
the resonance enhancement takes place at $m_\chi \sim m_H/2$. 
Operators $\mos_{3,4,5,6}$ couple to quarks only and so the branching ratios into quarks are similar, except for $\mos_6$
whose cross-section is proportional to the final state mass and hence dominated by the $\bar{b}b$ channel.

All in all, beside operator $\mos_5$, the rest of dim-6 operators cannot 
explain the antiproton excess since either the EFT approach breaks down~($\mos_{2, 4}$) or they are excluded
by the DD constraints~($\mos_{1,2,3,6}$). 
In fact, unlike the dim-5 operators above, some of the best-fit regions are quite close to the DD lower bounds and
therefore the DD limits can potentially be avoided by a moderate boost on the annihilation cross-section
from the underlying UV theory.

\section{Summary and Conclusion}

The antiproton flux excess  on the AMS02 data can be realized by DM annihilations into SM particles.  
Moreover, the DM mass range required to account for the excess   coincidentally overlaps with  those of  the DM explanation for the Galactic center gamma-ray excess~\cite{Hooper:2010mq,Hooper:2011ti}.
One, however, should carefully exam if other experimental bounds will rule out this possibility.
As a consequence, in the EFT framework we apply the constraints derived from 
the latest DM direct detection data to DM models
in the parameter space of the DM mass $m_\chi$ and the cut-off scale $\Lambda$.  
In this work, we study fermion DM only and focus on dim-5 and dim-6 operators listed in Eqs.~\eqref{eq:dim-5} and \eqref{eq:dim-6}. 

We first revisit the DM explanation for the AMS02 antiproton excess. 
The uncertainties of the DM signal from the propagation parameters 
and the background are properly included in the likelihood calculation 
based on a Bayesian approach. The likelihood calculation is further 
simplified with an approximation of the DM signal uncertainty 
calculation. That is, instead of computing the DM-induced antiproton 
flux for each set of the propagation parameters, we adopt a normalization
parameter $f$ to account for the uncertainty band of the DM signal. 
With this approximation we reproduces the actual likelihood distribution 
of the DM-induced antiproton component reasonably well, and the resulting 
best-fit parameter regions of DM are consistent with those obtained from 
the rigorous treatment. On the plane of the DM mass and the annihilation 
cross-section, we present the updated $95\%$ credible regions favored by 
the AMS02 antiproton data in Fig.~\ref{fig:aplike} for different final states.

Furthermore, we calculated the most updated DM direct detection likelihoods from 
LUX, PandaX-II, and XENON1T. 
From the likelihoods of the AMS02 antiproton and latest DD data, we attain the favored regions and exclusion limits respectively
on the plane of ($\mchi , \Lambda$).
In the DD analysis, the RG evolution of the Wilson coefficients and the matching to the nuclear theory are properly
taken into account.

 Note that for those operators with a velocity-suppressed annihilation cross-section, to realize the antiproton excess
the corresponding cut-off becomes so small that the EFT approach is no longer valid.     
On the other hand, the DM-nuclei scattering can be reliably computed based on EFT as the momentum exchange scale is
of $\mathcal{O}(100 \, \text{MeV})$, much smaller than the scale of interest. 

 We have found that the only effective operators which can reproduce the antiproton excess and avoid the DD bounds
 are those with an unsuppressed annihilation cross-section~($s$-wave) but with a suppressed~(either by velocity or momentum) DM-nuclei scattering interaction. Therefore, only $\mof_2$ and $\mos_5$ satisfy these two requirements.
 The former has a momentum-suppressed SI interaction~($\mathcal{Q}_{11}$) while the latter has a momentum-suppressed SD interaction~($\mathcal{Q}_{7,9}$). Besides, for dim-6 operators $\mos_{3,4,5,6}$, one can choose
 different couplings for the up- and down-type quarks to obtain ISV, alleviating the stringent DD bounds.   
 In this case, the antiproton best-fit regions of $\mos_{3,6}$ are actually not far below the DD limit in case of ISV. 
 That implies that if the underlying UV theory exhibits a certain boost mechanism for the annihilation cross-section   
 such as resonance enhancement, these kinds of operators can also accommodate the antiproton excess without being excluded
 by the DD experiments. On the other hand, operators $\mof_{1,3,4}$ have the best-fit regions far below the DD exclusion limit,
 requiring an enormous cross-section increase in the underlying UV model which becomes less natural.

 To summarize, operators $\mof_2$ and $\mos_5$ can successfully realize the antiproton excess without being excluded
 by the DD experiments,  LUX, PandaX-II, and XENON1T. For some of the operators such as $\mos_{3,6}$, the best-fit region
 is close to the DD limit in case of ISV.
 The underlying model with a moderate boost on the annihilation cross-section can still reproduce the excess
 and escape the DD constraints.

\acknowledgments

This work is supported by the National Key Research and Development Program 
of China (No. 2016YFA0400204), the National Natural Science Foundation of 
China (No. 11722328), and the 100 Talents program of Chinese Academy of 
Sciences.
WCH is supported by the Independent Research Fund Denmark, grant number 
DFF 6108-00623.

\appendix

\section{Comparison between results of the exact and approximate likelihoods for the antiproton flux}
\label{sec:appA}

\begin{figure}[!htb]
\includegraphics[width=0.7\textwidth]{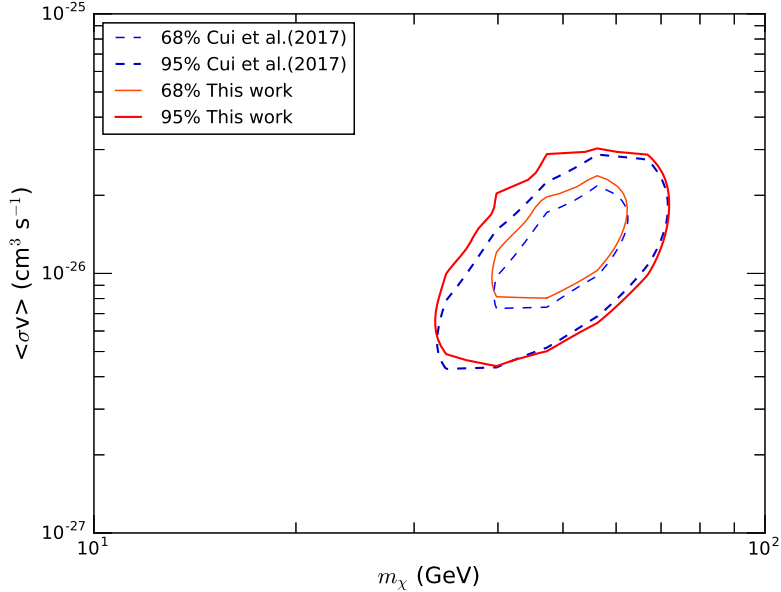}
\caption{The 68\% and 95\% credible regions on the DM parameter plane 
$(m_{\chi},\sv)$ favored by the AMS-02 antiproton data, assuming the $b\bar{b}$ channel. 
\label{fig:compare}}
\end{figure}

In Fig.~\ref{fig:compare}, we compare the constraints on the
DM model parameters $(m_{\chi},\sv)$ between the exact likelihood
calculation using Eq.~(\ref{eq:post_dm}) and the approximate likelihood
using Eq.~(\ref{eq:post_dm2}). It is clear that the approximate 
likelihood is able to reproduce the exact result quite well. 

\section{The annihilation cross-sections for $\mof_3$ and $\mof_4$}
\label{sec:appB}

In this work, we utilize the package \texttt{FeynRules}~\cite{Alloul:2013bka} to generate the interaction vertices 
that then are imported into \texttt{MicrOMEGAs}~\cite{Barducci:2016pcb} to compute the corresponding annihilation cross-sections. 
For $\mof_3$ and $\mof_4$, however, we manually computed the annihilation cross-section and the results are:     
\begin{eqnarray}
   \sv_{3}^{(5)}&=&\sum_f\frac{Q_f^2 }{48 \pi  \Lambda^2} 
                   \frac{m_f^4\left(13 v^2-24\right)+4 m_f^2 \mchi^2 \left(v^2-6\right)-
                   8 \mchi^4
                   \left(v^2-6\right)}{ \mchi^3 \sqrt{\mchi^2-m_f^2}}, 
\\
\sv_{4}^{(5)}&=&\sum_f\frac{Q_f^2 }{24 \pi \Lambda^2}
               \frac{v^2 \sqrt{\mchi^2-m_f^2} \left(m_f^2+2 \mchi^2\right)}
               { \mchi^3},   
\label{eq:ann5}
\end{eqnarray}
where the parameters $Q_f$ and $m_f$ are the electric charge and mass of the SM fermion $f$, respectively. 
We set the relative velocity to be $v\sim 10^{-3} \, c$.

\section{DM non-relativistic operators for direct detection}
\label{sec:appC}
The non-relativistic quantum mechanical operators are listed in the Table~\ref{tab:operators}, 
obtained from Refs.~\citep{Fitzpatrick:2012ix,Anand:2013yka}. 
The operator ${{\bf{S}}}_{\chi}$~(${{\bf{S}}}_{N}$) is the DM~(nucleon) spin,
the vector $\bf{{q}}$ is the transfer momentum, and 
${\bf{{v}}}^{\perp}$ is the velocity perpendicular to $\bf{{q}}$,
defined as ${\bf{{v}}}^{\perp}={\bf v}+{\bf q}/(2\mu_{\chi N})$ where 
$\mu_{\chi N}$ is the DM-nucleon reduced mass.
The operators from $\mathcal{O}_1$ to $\mathcal{O}_{11}$ correspond to  
interactions mediated by spin-0 or spin-1 particles.

\begin{table}[t]
    \centering
    \begin{tabular}{ll}
    \hline\hline
         {${\mathcal{Q}}_1 = \textbf{1}_{\chi}\textbf{1}_{N}$}
         & ${\mathcal{Q}}_9 = i{\bf{{S}}}_\chi\cdot\left({{\bf{S}}}_N\times\frac{{\bf{{q}}}}{m_N}\right)$  \\
        ${\mathcal{Q}}_3 = i{{\bf{S}}}_N\cdot\left(\frac{{\bf{{q}}}}{m_N}\times{\bf{{v}}}^{\perp}\right)$ \hspace{2 cm} &   ${\mathcal{Q}}_{10} = i{{\bf{S}}}_N\cdot\frac{{\bf{{q}}}}{m_N}$   \\
        ${\mathcal{Q}}_4 = {{\bf{S}}}_{\chi}\cdot {{\bf{S}}}_{N}$ &   ${\mathcal{Q}}_{11} = i{\bf{{S}}}_\chi\cdot\frac{{\bf{{q}}}}{m_N}$   \\                                                                             
        ${\mathcal{Q}}_5 = i{\bf{{S}}}_\chi\cdot\left(\frac{{\bf{{q}}}}{m_N}\times{\bf{{v}}}^{\perp}\right)$ &  ${\mathcal{Q}}_{12} = {{\bf{S}}}_{\chi}\cdot \left({{\bf{S}}}_{N} \times{\bf{{v}}}^{\perp} \right)$ \\                                                                                                                 
        ${\mathcal{Q}}_6 = \left({\bf{{S}}}_\chi\cdot\frac{{\bf{{q}}}}{m_N}\right) \left({{\bf{S}}}_N\cdot\frac{{{\bf{q}}}}{m_N}\right)$ &  ${\mathcal{Q}}_{13} =i \left({{\bf{S}}}_{\chi}\cdot {\bf{{v}}}^{\perp}\right)\left({{\bf{S}}}_{N}\cdot \frac{{\bf{{q}}}}{m_N}\right)$ \\   
        ${\mathcal{Q}}_7 = {{\bf{S}}}_{N}\cdot {\bf{{v}}}^{\perp}$ &  ${\mathcal{Q}}_{14} = i\left({{\bf{S}}}_{\chi}\cdot \frac{{\bf{{q}}}}{m_N}\right)\left({{\bf{S}}}_{N}\cdot {\bf{{v}}}^{\perp}\right)$  \\
        ${\mathcal{Q}}_8 = {{\bf{S}}}_{\chi}\cdot {\bf{{v}}}^{\perp}$  & ${\mathcal{Q}}_{15} = -\left({{\bf{S}}}_{\chi}\cdot \frac{{\bf{{q}}}}{m_N}\right)\left[ \left({{\bf{S}}}_{N}\times {\bf{{v}}}^{\perp} \right) \cdot \frac{{\bf{{q}}}}{m_N}\right] $ \\                                                                               
    \hline\hline
    \end{tabular}
    \caption{Non-relativistic effective operators, taken from Ref.~\citep{Anand:2013yka}, describing
    DM-nucleon interactions.} 
    \label{tab:operators}
\end{table}

\end{document}